\documentstyle[12pt]{article}
\def\beq{\begin{equation}}
\def\eeq{\end{equation}}
\def\bea{\begin{eqnarray}}
\def\eea{\end{eqnarray}}
\def\bq{\begin{quote}}
\def\eq{\end{quote}}

\def\gappeq{\mathrel{\rlap {\raise.5ex\hbox{$>$}}
{\lower.5ex\hbox{$\sim$}}}}

\def\lappeq{\mathrel{\rlap{\raise.5ex\hbox{$<$}}
{\lower.5ex\hbox{$\sim$}}}}

\def\Toprel#1\over#2{\mathrel{\mathop{#2}\limits^{#1}}}

\begin{document}
\pagestyle{empty}
\begin{flushright}
{CERN-TH/2000-148} \\
IITAP-2000-010 \\
hep-ph/0005279 \\
\end{flushright}
\vspace*{5mm}
\begin{center}
{\Large \bf Parton Scattering at Small-$x$ and Scaling 
Violation}${}^{\star}$ \\
\vspace*{1cm} 
{\large  \bf Victor~T.~Kim${}^{\& \dagger }$,
Grigorii~B.~Pivovarov${}^{\S }$
{\rm and}
James~P.~Vary${}^{\$ \ddagger}$}
\end{center}
${}^\& $ : CERN, CH-1211, Geneva 23, Switzerland
\newline
${}^\S$ : Institute for Nuclear Research, 117312 Moscow, Russia
\newline
${}^\$ $ : Int. Inst. of Theoretical and Applied Physics,
Iowa State University, Ames,
IA 50011, USA
\newline
${}^\ddagger$ : Department of Physics and Astronomy,
Iowa State University, Ames,
IA 50011, USA
\vspace*{2cm}
\begin{center}  
{\bf ABSTRACT} \\ \end{center}
\vspace*{5mm}
Scaling violation of inclusive jet production at small-$x$ in hadron
scattering  with increasing total collision energy
is discussed. Perturbative QCD based on the factorisation
theorem for hard processes and GLAPD evolution equations
predicts a minimum for scaled cross-section ratio
that depends on jet rapidity.
Studies of such a scaling violation can reveal a vivid indication
of new dynamical effects in
the high-energy limit of QCD. The BFKL effects, which seem to be seen
in recent L3 data at CERN LEP2, should give  different results
from GLAPD predictions.

\vspace*{2cm} 
\noindent 
\rule[.1in]{16.5cm}{.002in}

\noindent
${}^{\star}$ Based on the talk by VTK at
the Xth Quantum Field Theory and High Energy Physics Workshop (QFTHEP'99),
Moscow, Russia, May 27 - June 2, 1999, to appear in the Proceedings.

\noindent
$^{\dagger}$ Permanent address: 
 St.Petersburg Nuclear Physics Institute,
188300 Gatchina, Russia; \\
e-mail: kim@pnpi.spb.ru. 
\vspace*{0.5cm}

\begin{flushleft} CERN-TH/2000-148 \\
April 2000
\end{flushleft}
\vfill\eject

\setcounter{page}{1}
\pagestyle{plain}

QCD is an essential ingredient of the Standard Model, 
and it is well tested
in  hard processes when transferred momentum is of the order
of the
total collision energy (Bjorken limit: $Q^2 \sim s \rightarrow \infty$).
The cornerstones of perturbative QCD
at this kinematic regime (QCD-improved parton model):
factorization of inclusive hard processes \cite{Amati}
and the Gribov--Lipatov--Altarelli--Parisi--Dokshitzer
(GLAPD) evolution equation \cite{GLAPD} provides
a basis for the successful
QCD-improved parton model. The factorisation theorem  \cite{Amati}
for inclusive hard  processes
ensures that the inclusive cross section  factorises
into partonic subprocess(es)
and parton distribution function(s).
The GLAPD evolution equation governs the $\log Q^2$-dependence
(at $Q^2 \rightarrow \infty$)
of the
inclusive hard process cross-sections at fixed scaling variable $x=Q^2/s$.

Another kinematic domain that is very important
at high-energy is given by the
(Balitsky--Fadin--Kuraev--Lipatov)
BFKL limit 
[3--6], 
or QCD Regge limit,
whereby at fixed $Q^2 \gg \Lambda_{QCD}^2$, $s \rightarrow \infty $.
In the BFKL limit, the BFKL evolution in the leading order (LO)
governs $\log(1/x)$ evolution (at $x \rightarrow 0$)
of inclusive processes.
Note that the BFKL evolution
in the next-to-leading order (NLO) 
[7--10],
unlike the LO BFKL 
[3--5],
partly includes GLAPD evolution
with the running coupling constant of the LO GLAPD,
$\alpha_S(Q^2) = 4 \pi / \beta_0
\log(Q^2/\Lambda_{QCD}^2)  $.

Therefore, the BFKL and especially the NLO BFKL 
[7--10]
are anticipated to be
important tools for exploring the high-energy limit of QCD.
In particular, this importance
arises since the highest eigenvalue, $\omega^{max}$, of the BFKL
equation 
[3--6, 9, 10]
is related to the intercept of the Pomeron, which in turn governs
the high-energy asymptotics of the total cross-sections: $\sigma \sim
(s/s_0)^{\alpha_{I \negthinspace P}-1} = (s/s_0)^{\omega^{max}}$, where
the Regge parameter $s_0$ defines the approach to
the asymptotic regime.
The BFKL Pomeron intercept in the LO turns out to be rather large:
$\alpha_{I \negthinspace P} - 1 =\omega_{LO}^{max} =
12 \, \log2 \, ( \alpha_S/\pi )  \simeq 0.54 $ for
$\alpha_S=0.2$; hence, it is very important to analyse
recently calculated  NLO corrections \cite{FL,KL} to the BFKL.

One of the striking features of the NLO BFKL analysis \cite{BFKLP}
is that the NLO value for
the intercept of the BFKL Pomeron, improved by the BLM procedure \cite{BLM},
has a very weak dependence on the gluon virtuality $Q^2$:
$\alpha_{I \negthinspace P} - 1 =\omega_{NLO}^{max}   \simeq$ 0.13 -- 0.18 
at $Q^2 = 1$ -- 100 GeV$^2$.
This agrees with the conventional Regge theory where
one expects  universal intercept of the Pomeron without any $Q^2$-dependence.
The minor $Q^2$-dependence obtained leads
to approximate conformal invariance.

There have recently been a number of papers which analyse the NLO BFKL
predictions
[12--16].
Also, a lot of work should be 
done to clarify the very important issue
of the factorisation properties of the BFKL regime
[17--23].

\begin{figure}[htb]
\vskip 7 cm
\includegraphics{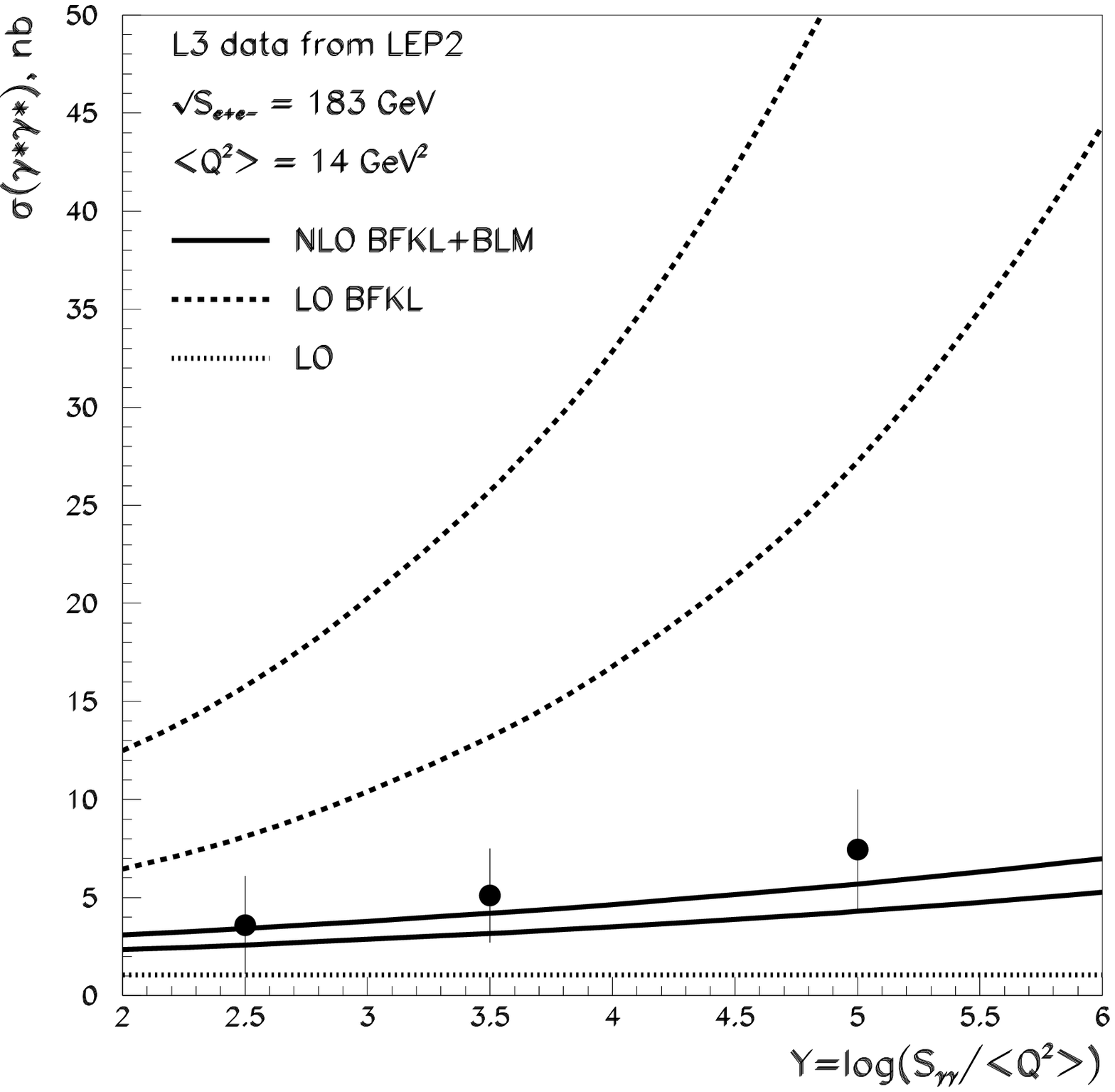}
\vskip 0 cm
\includegraphics{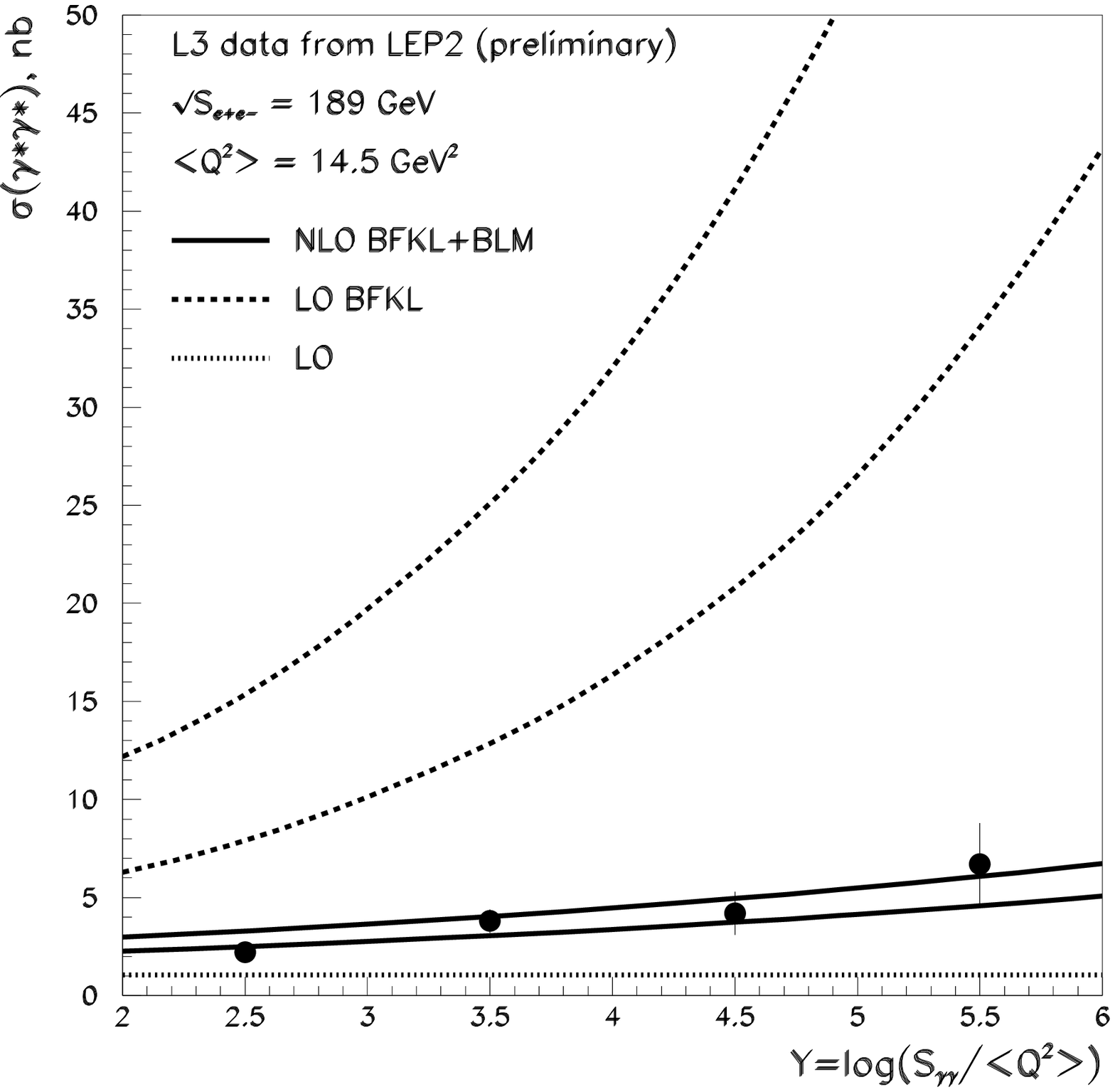}
\caption{
Virtual gamma--gamma total cross-section
by the BFKL Pomeron versus 
L3 Collaboration data at energies: (a) 183 GeV and 
(b) 189 GeV of $e^{+}e^{-}$ collisions.
Solid curves correspond to NLO BFKL in BLM; dashed: LO BFKL;
and dotted: LO Born contribution. Two different curves are for
two different choices of the Regge scale: $s_0= Q^2/2$ and $s_0=2 Q^2$.}
\label{fig:1}
\end{figure}

As a phenomenological application of the NLO BFKL
improved by the BLM procedure,
with its effective resummation of
the conformal-violating
$\beta_0$-terms into the running coupling in all orders
of the perturbation theory, one can consider
the gamma--gamma scattering \cite{BFKLP00,KLP}.
This process is attractive because
it is theoretically more under control than the hadron--hadron
and lepton--hadron collisions, where nonperturbative hadronic structure
functions are involved. In addition, for the gamma--gamma scattering
the unitarisation (screening) corrections
due to multiple Pomeron exchange would be less important than in
hadron collisions.

The gamma--gamma cross sections with the BFKL resummation in the LO
were considered in \cite{BL78,Brodsky97,Bartels96}.
In the NLO BFKL case one should obtain a formula analogous to
LO BFKL \cite{BFKLP00}.

In Fig. \ref{fig:1} we present the comparison of
BFKL predictions for LO and NLO
BFKL improved by the BLM procedure with data
\cite{L3} from L3 at CERN LEP. The different curves reflect
the uncertainty of the theoretical predictions with the
choice of the Regge scale parameter $s_0$, which defines
the transition to the asymptotic regime.
For the present calculations two variants have been chosen $s_0=Q^2 / 2$
and $s_0 = 2 Q^2$, where $Q^2$ is the virtuality of the photons.
One can see from Fig. \ref{fig:1}
that the agreement of the NLO BFKL improved by the BLM procedure
is reasonably good at energies of LEP2
$\sqrt{s_{e^+e^-}}=$ 183 -- 189 GeV.
One can notice also that the sensitivity of the NLO BFKL results
to  $s_0$ is much smaller than in
the case of the LO BFKL.

 It was shown in Refs. \cite{Dubovikov77,Kaidalov86}
that the unitarisation corrections in hadron collisions
can lead to a value of the (bare) Pomeron intercept higher than
the effective intercept value. Since the hadronic data fit yields
about 1.1 for the effective intercept value \cite{Cudell99},
the bare Pomeron intercept value should be above it.
Therefore, assuming small unitarisation corrections in the gamma--gamma
scattering at large $Q^2$, one can accommodate the NLO BFKL Pomeron
intercept value 1.13 -- 1.18 \cite{BFKLP} in the BLM
optimal scale setting, along with larger unitarisation corrections
in hadronic scattering \cite{Kaidalov86,Cudell99}, where they can lead to
a smaller effective Pomeron intercept value of about 1.1 for
hadronic collisions. The above intercept value of the NLO BFKL 
Pomeron is in good agreement with  the
analysis \cite{Cox99} of the diffractive 
dijet production at the Tevatron.

Another possible application of the BFKL approach
can be the collision energy dependence of the {\it inclusive}
jet production \cite{KP98,KP97}.
Unlike the case with the selection of most forward/backward
(Mueller-Navelet) jets  
[19, 33--38],
the usual {\it inclusive} jets 
[20--22, 31, 39, 40]
can be more reliable for detectors with the limited
acceptance in rapidity.

The advent of the Fermilab Tevatron and the CERN LHC
provides a new testing ground for the parton
model---the kinematic conditions when the energies of the produced hadrons
are large enough to be described by
perturbation theory and,  at the same time, are much smaller than the total
energy of the collision (BFKL semi-hard kinematics).
Because the parton model was originally invented and subsequently tested for
the hard kinematics, the second condition makes it plausible
that a substantial modification of the parton model will be needed to describe
this BFKL semi-hard kinematic region.

The range of applicability of the QCD-improved parton model
is a subject of controversy at the moment. There are statements
(see, {\it e.g.}, \cite{Bal94,Ynd96}) that the fitting capacity
of the conventional QCD-improved parton model is sufficient
to accommodate all the data on deep inelastic scattering (DIS)
parton structure functions available
at small-$x$ kinematics.
On the other hand, the same data from HERA on DIS structure functions can be
described by NLO BFKL \cite{Thorne,Altarelli}. In addition,
the data from HERA \cite{HERA98} and Tevatron \cite{D099} on most forward/backward
jet production  may be interpreted as a manifestation of the BFKL
Pomeron \cite{Bar96,MN}, which is beyond
the conventional QCD-improved parton model.
The situation is further complicated by the observation that
the range of applicability of the QCD-improved parton model
may be different
for different observables. In particular, the cross-sections of
processes with specific kinematics exhibit
breakdown of the applicability of finite-order
perturbative QCD via the 
development of sensitivity to the choice of
the normalisation scale. On the other hand, some dedicated combinations
(ratios) of cross sections may be less sensitive to the inclusion of the
higher-order corrections. An example is the scaled cross-section
ratio 
[45--48]
since, as 
follows from  Ref. \cite{Ell93}, it is relatively
insensitive to the inclusion of the NLO correction.

Under these circumstances, it is crucial to have {\it qualitative}
predictions from the conventional QCD-improved parton model (without
BFKL resummation of the energy logarithms) for the new kinematic domain. If the
predictions would turn out qualitatively incorrect, a generalisation,
and a substitute in this kinematical domain
of the QCD-improved parton model would become indispensable.

\begin{figure}[htb] 
\vskip 7 cm
\includegraphics{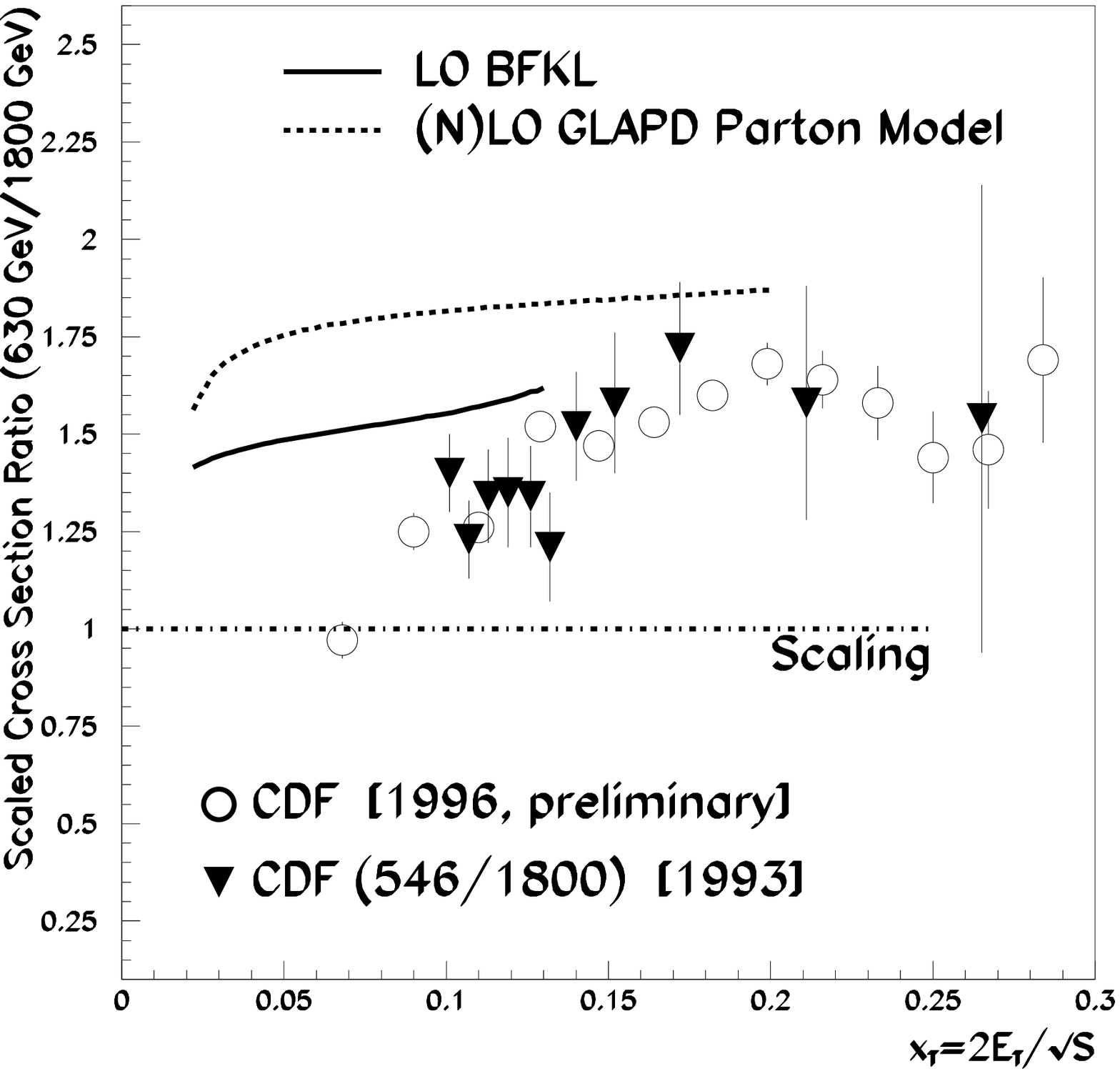}
\vskip 0 cm
\includegraphics{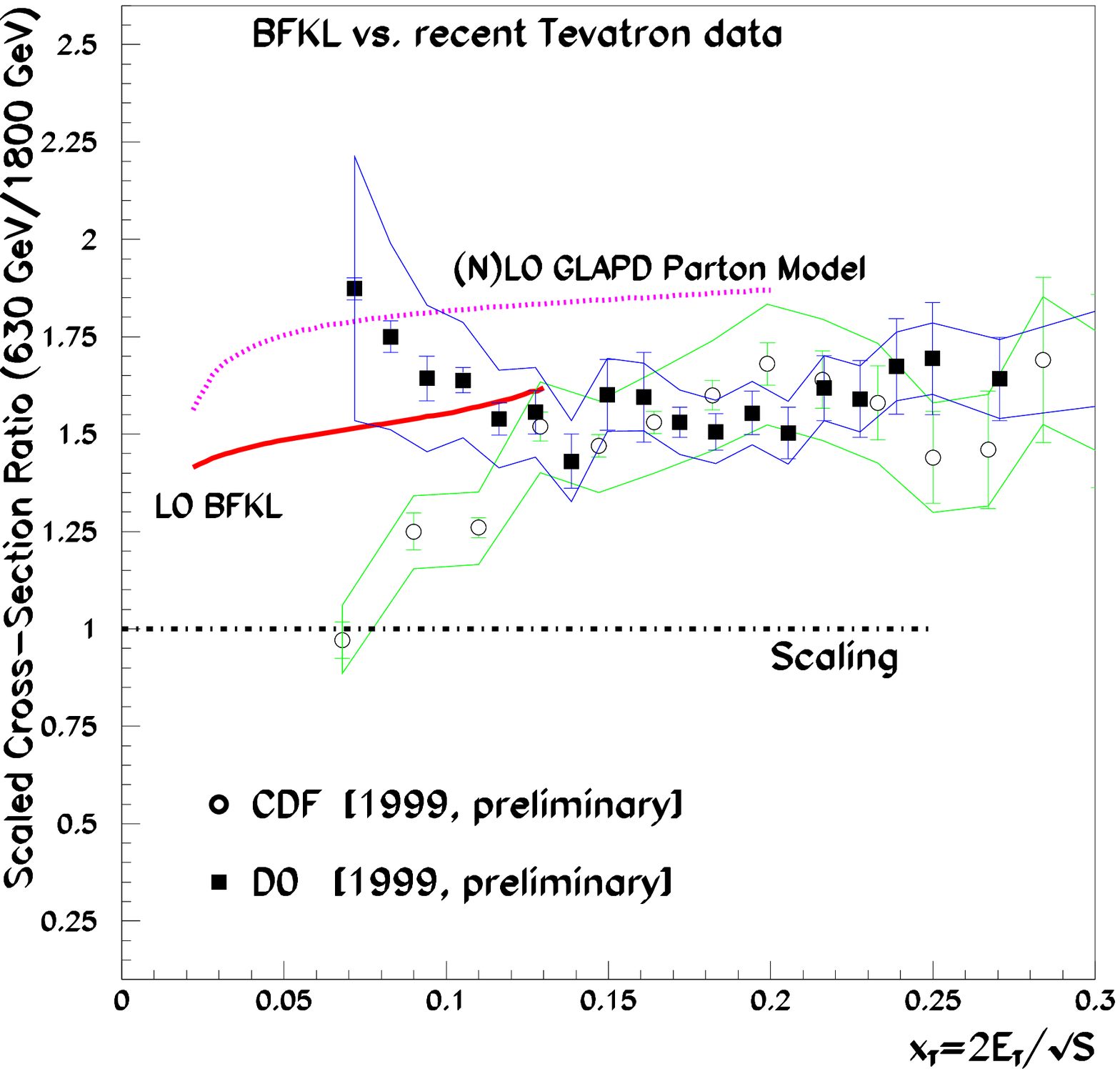}
\caption{
Inclusive single-jet production: 
scaled cross-section ratio at the Tevatron. 
BFKL \cite{KP98,KP97} and GLAPD \cite{Ell93} predictions
versus Tevatron data: (a) CDF data \cite{CDF93,CDF96}; 
(b) preliminary data from  CDF and D$\emptyset$ 
Collaborations \cite{Hus98}. }
\label{fig:2}
\end{figure}

Here we discuss such a prediction, made in Ref. \cite{KPV}. It is a prediction
for the ratio of inclusive single jet production at a smaller energy
$\sqrt{s_N}$
of the hadron collision to the one at a higher energy $\sqrt{s_D}$:
\begin{equation}
\label{Rdef}
R(x, y, s_N, s_D) =  \frac{s_N d\sigma}{dxdy}(x,y,s_N) \Big/
\frac{s_D d\sigma}{dxdy}(x,y,s_D) .
\end{equation}
Here the cross-section is made dimensionless by the rescaling with the
corresponding total invariant energy of the collision squared
$s_N (s_D)$.
The ratio depends on the (pseudo)rapidity $y=1 / 2 \ln (k_+ / k_-)$,
where $k_\pm = E \pm k_3$ are
the light-cone components  of the momentum of the produced jet, and on the
fraction of the energy
$x = (k_+ + k_-) / \sqrt{s_i}$, $i=N,D$ deposited in the jet produced
($s_N$ is used for the definition of $x$ in the numerator, $s_D$ in the
denominator, so that $x$ varies from zero to unity for both energies).
Note that this scaling variable coincides in the centre-of-mass system
with  $x_R = E/E_{max} = 2 E/ \sqrt{s}$ ,
the radial Feynman variable, or for $y=0$, {\it i.e.}
for $(\theta_{CMS}=\pi/2)$,
the scaling variable becomes the transverse Feynman variable:
$x=x_{\perp}=2 E_\perp / \sqrt{s}$.

The ratio $R$ (taken at $y \simeq 0$, {\it i.e.} for jets perpendicular
to the collision axes) was used in Refs. \cite{UA285,CDF93,CDF96,Hus98}
as a means to test QCD predictions
for scaling violations:

\begin{eqnarray}
\label{Rperp}
R(x, y=0, s_N, s_D) &= &
E_{\perp}^4 \frac{ E d\sigma}{d^3 p}(x_\perp,y=0,s_N) \Big/
E_{\perp}^4\frac{ E d\sigma}{d^3 p} (x_\perp,y=0,s_D) \\
   & = &
E_{\perp}^4 \frac{ d\sigma}{dy d E_\perp^2} (x_\perp,y=0,s_N)\Big/
E_{\perp}^4 \frac{ d\sigma}{dy d E_\perp^2} (x_\perp,y=0,s_D).
\end{eqnarray}

Note that without scaling violations the
ratio $R$ is {\it exactly} unity. At fixed $y$ and $x_\perp$
the dependence of $R$ from $s_{N}$, $s_{D}$ comes from the presence of
the fundamental QCD scale, $\Lambda_{QCD}$, in the running coupling
and in the parton distribution functions.

To be more explicit, we remark that the QCD-improved parton model,
based on the factorisation theorem for hard processes and
the GLAPD $\log Q^2$-evolution, presents the inclusive jet
scaled cross-section in hadron collsions as

\begin{equation}
E_{\perp}^4 \frac {E d\sigma}{d^3 p}(x_\perp,y=0,s) =
 \int^{1}_{x_{A, min}} \int^{1}_{x_{B, min}} dx_A dx_B
 F_A(x_A,Q^2) F_B(x_B,Q^2) E_{\perp}^4 \frac{\hat{s}}{\pi}
 \frac{d \hat{\sigma}} {d \hat{t}} \, \delta (\hat{s} + \hat{t} + \hat{u}),
\end{equation}
where
$\hat{s}$, $\hat{t}$ and $\hat{u}$ are the Mandelstam variables for the
partonic subprocess, the scale of the hard partonic subprocess
$ - \hat{t} = Q^2 \sim E_\perp^2 \sim x_\perp^2 s$,
$F_A$ and $F_B$ are parton distribution functions with the
GLAPD evolution following from perturbative $\alpha_S(Q^2)$ expansion,
and the scaled partonic subprocess is 

$$ E_{\perp}^4 \frac{d \hat{\sigma}} {d \hat{t}} \sim
 \alpha_{S}^2(Q^2) [1+ C_{NLO} \alpha_{S} + ...]
= \alpha_{S}^2 (x_{\perp} s^2) [1+ C_{NLO} \alpha_{S} + ...] . $$

Hence, within the QCD-improved parton model, the scaled
cross-section ratio for inclusive jet production  at fixed $x$
and $y$  is the dimensionless function of $\alpha_S$.
The GLAPD scaling violation due to the interacting QCD partons
appears as the {\it logarithmic}\footnotemark
\footnotetext{Indeed,
the small-$x$ asymptotics of
the GLAPD evolution gives a growth of parton structure functions
that is faster than any power of a logarithm, but slower
than any power --- so-called double-logarithmic asymptotics
\cite{Ruj74,GLR,Olness,Bal94}.}
dependence on the total energy of collision
through the coupling constant $\alpha_S$.
We note here that the BFKL leads to a {\it power-like}
scaling violation, the strength of which depends on the Regge scale $s_0$.

Perturbative QCD calculations of Ref. \cite{Ell93} with hard kinematics
($Q^2  \sim  s $) predict for $R$ at $y=0$
a steep increase around the value of 1.8 -- 1.9 for $x$ growing in the
range above 0.1  (for the case $\sqrt{s_N}$/$\sqrt{s_D}$ =
0.63 TeV/1.8 TeV ).   For moderate $x$, the prediction is in reasonable
agreement with CDF data  \cite{CDF93,CDF96}.
For $x<0.1$, calculations are above the preliminary data of CDF \cite{CDF96}.
This was one of the reasons for the conclusion of  Ref.
\cite{Ell93} that
NLO GLAPD \cite{Ell90} with hard kinematics
is insufficient to describe the absolute cross section
of jets with  transverse energy less  than 50 GeV within  an
accuracy of $10 \% $. It was shown in Ref. \cite{KP98} that
resummation of the energy
logarithms, {\it i.e.} BFKL, restores the agreement between theory and experiment.

In Ref. \cite{KPV}  we have found the following result:
the QCD-improved parton model predicts that $R$ is not a monotonic
function of its arguments, {\it i.e.}
the single-jet production cross-section, if
measured in the natural units of the same cross section taken at another
(higher) energy of the collision, has extrema. Namely, it has minima
(``dips''):
there is a value of $x$ for each $y$ with the smallest ratio of jets
produced. The reason this fact was overlooked is that for $y=0$
(the only value for which
the calculations were reported earlier) the minimum is at a
value of $x$ too small to be inside the acceptance of the existing detectors
($x_{dip}(y=0)<0.01$ at the Tevatron).

\begin{figure}[htb]
\vskip 7 cm
\includegraphics{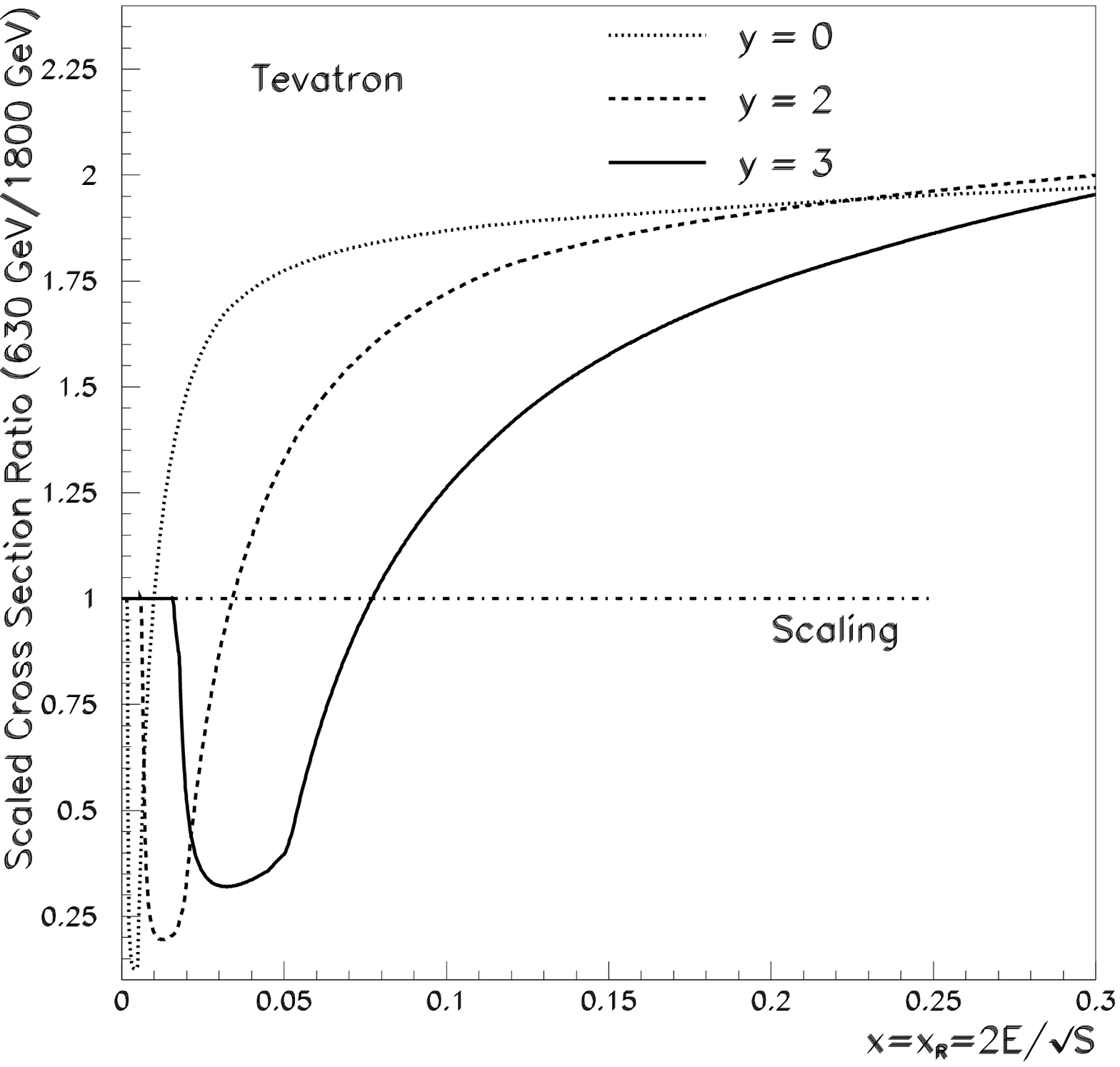}
\vskip 0 cm
\includegraphics{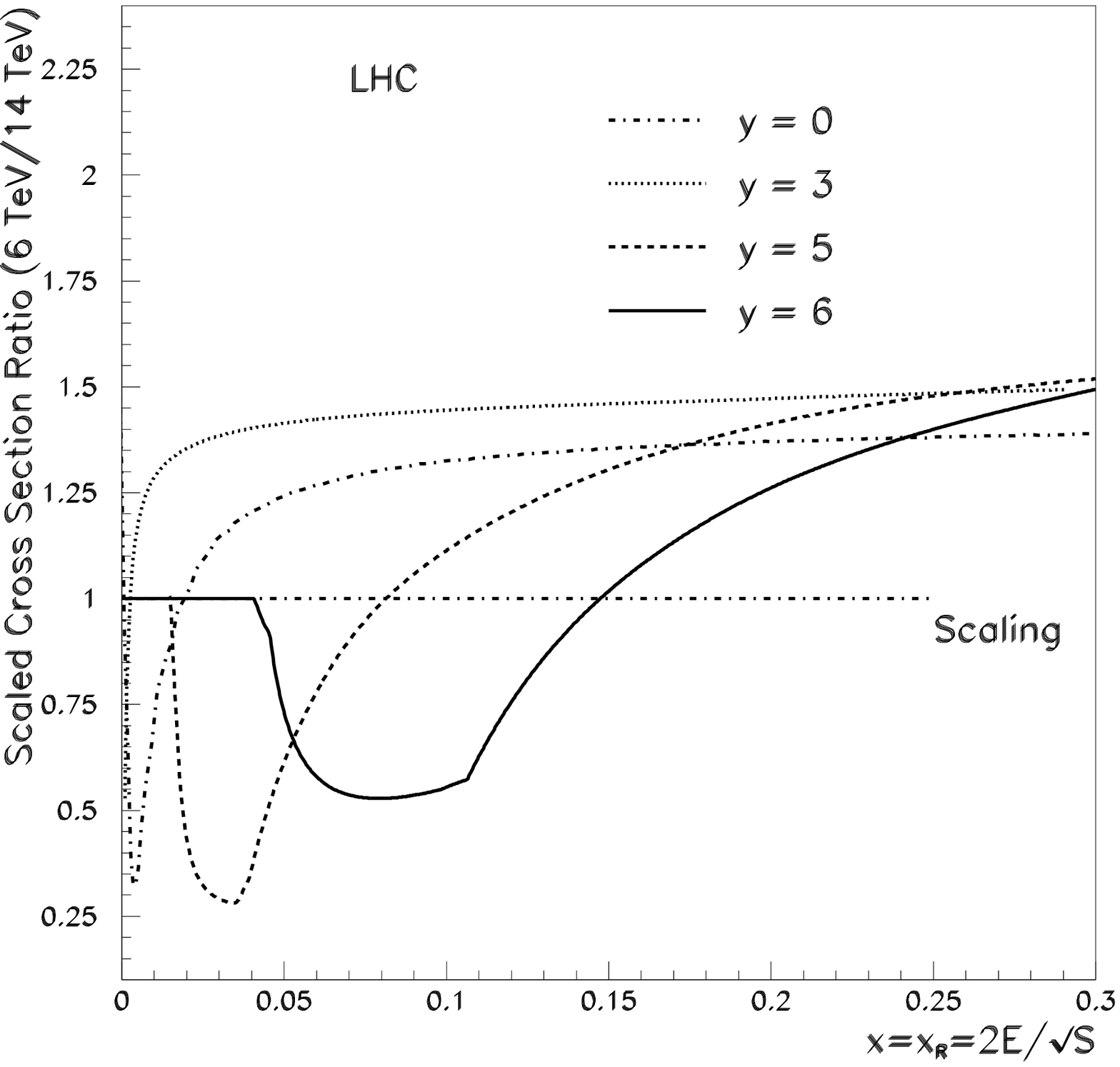}
\caption{Ratio of scaled cross-sections of inclusive 
single-jet production: (a) for Fermilab energies;
(b) for the LHC energies.}
\label{fig:3}
\end{figure}

Fig. \ref{fig:3}a presents the ratio for energies 0.63 TeV/1.8 TeV at the
Fermilab Tevatron, Fig. \ref{fig:3}b for energies 6 TeV/14 TeV at the
CERN LHC.
Each curve on the plots presents the dependence of the ratio
on $x$ at different values of the rapidity $y$.
Each curve ends at a lowest value of $x$ where
$\alpha_S(Q^2)$
has the value of about 0.5 (it corresponds
to $Q=0.7$ GeV, and $Q$ was taken to be half of the transverse
energy of the jet produced).
For lower values of $x$, perturbative theory becomes unreliable
because the coupling approaches unity.

There is another lower bound on the values of $x$ at which our plots
make sense, because there is a lowest energy for which the jet may
be resolved. This energy is accepted now to be around 5 GeV
\footnote{At the  HERA lepton--hadron collider jets are resolved
from $E_\perp=3$ GeV, and at the Tevatron hadron--hadron collider
the CDF Collaboration tags jets from
$E_\perp=8$ GeV
\cite{CDF98}. },
which corresponds to $x>0.016$ for the Tevatron (Fig. \ref{fig:3}a),
and to $x>0.0017$ for the LHC (Fig. \ref{fig:3}b).

There is an important issue concerning the accuracy of the present
LO calculation. The most important
advantage of the scaled cross-section ratio is
that this is the ratio of two perturbative series
with the same coefficients and with different scales
in the running coupling. These scales are defined
by the two initial collision energies at fixed
scaling variable. One can show that the theoretical
accuracy of the ratio in LO of perturbative QCD
is not less than the accuracy of NLO
calculations for absolute cross-sections.

The minima in Fig. \ref{fig:3} originate from a competition between the
running of the parton distribution functions and the running of the
coupling constant. Namely, the ratio with frosen parton
distribution functions is  decreasing monotonously (this tendency is
realised at small $x$),  while the one with frosen coupling  constant is
growing monotonously (which is realised for
$x$ larger than the position of the minimum).

We suggest the following potential implications of the minima we
have predicted with the parton model: (i) If one observes
the minima experimentally, one employs the orthodox QCD-improved
parton model and tries to account for observed positions
and depths  of the minima by taking into account higher-order
corrections, in particular, resummation of the energy logarithms.
(ii) If one does not observe the minima experimentally,
more radical changes are motivated, such as an
alternative model of the effective constituents
inside the hadrons for the BFKL semi-hard asymptotics.
One example might be the colour dipole model \cite{Mue94}.

Finally, we comment on the possibility of searching
for the minima at the Fermilab  Tevatron and at the CERN LHC:
positions of the minima for  the Tevatron energies
(Fig. \ref{fig:3}a) seem to be reached
by both D$\emptyset$ and CDF detectors.
The minima of the LHC plot
(Fig. \ref{fig:3}b) seem to be well inside the acceptance of, 
for example,
the FELIX \cite{Felix}, the ALICE \cite{Alice} and
the CMS \cite{CMS} detectors.

We take the ratio of 6 TeV/14 TeV for the LHC,
because, in addition to 14 TeV $pp$ collisions,
lead--lead collisions at the LHC are planned with
the collision energy  of 6 TeV per nucleon--nucleon collision.
Since nuclear collisions bring in nuclear effects, which can distort
our predicted curves, we also considered the ratio 6 TeV/100
TeV (see Ref. \cite{KPV}).

Further consideration should be given to deciding
which pair of energies and  value of rapidity are  most
convenient for an experimental search of the minima.
Also, more work is needed to make quantitative predictions
on the locations and the shapes of the dips with the NLO
corrections taken into account. It is interesting to observe a resemblance
of the dips presented here with the nonmonotonic behaviour of
parton distribution functions in DIS \cite{Jenk99}.

It is also worth noting that, in the case of nuclear collisions,
the effects of initial nuclear parton distributions (small-$x$ EMC--effect
\cite{EMC}) and dynamical effects such as quark--gluon plasma,
jet quenching, etc. \cite{Wang97}, phase structure of the QCD
vacuum \cite{tHooft,KMPV} will demand special consideration.

Before reaching conclusions, we would like to note that many of
the above ideas can be studied also for the case of heavy-quarkonium
production, where similar phenomena should be present
\cite{KPSV}.

To sum up, we find a new qualitative prediction of the
QCD-improved parton model for hadron collisions and
suggest its use to test the applicability of the
parton model for certain regions of high-energy
hadron collisions. Study of the scaling violation
of the scaled cross section ratio can reveal such new
dynamic effects as the BFKL asymptotics.

We thank G. Altarelli, V. P. Andreev, J. Ellis, A. De Roeck, 
V. S. Fadin, J. H. Field, M. Kienzle-Focacci, C.-H. Lin, L. N. Lipatov, J.-W. Qiu, 
V. A. Schegelsky, A. A. Vorobyov and M. Wadhwa for useful discussions,
 and, also, S. Vascotto for reading of the manuscript.
VTK thanks the Organizing Committee of the Xth Quantum Field Theory and
High Energy Physics  Workshop  and the CERN Theory Division for 
their warm hospitality.  This work was supported in part by
the Russian Foundation for Basic Research, Grant No. 00-02-17432,
the NATO Science Programme, Collaborative Linkage Grant No. PST.CLG.976521, 
and the U.S. Department of Energy, Contract No. DE-FG02-87ER40371,
Division of High Energy and Nuclear Physics.

\end{document}